# Original Articles

## Promising outcomes of an online course in research writing at a Rwandan university

**Ravi Murugesan**
*AuthorAID Training Coordinator, International Network for the Availability of Scientific Publications (INASP), 60 St Aldates, Oxford OX1 1ST, UK; ravi@uwalumni.com*

**Abstract**

*Background:* Researchers in developing countries often do not have access to training on research writing. The purpose of this study was to test whether researchers in Rwanda might complete and benefit from a pilot online course in research writing.

*Methods:* The pilot course was set up on Moodle, an open-source online learning environment, and facilitated by the author. The lessons and assignment were spread over six weeks, followed by a two-week extension period. Twenty-eight faculty members of the National University of Rwanda enrolled themselves in the course.

*Results:* Twenty-five of the 28 learners completed the course. After the course, these learners expressed high satisfaction, eg, 24 of them felt that they were ready to write a research paper for publication.

*Conclusion:* The high completion rate (89%) is noteworthy for two reasons: e-learning courses tend to have lower completion rates than classroom courses, and 76% of the learners in the pilot course had not taken an e-learning course before. This result and the positive feedback indicate that online courses can benefit researchers in developing countries who may not have access to classroom courses on research writing.

## Introduction

*Background*

The under-representation of research publications from developing countries has caused concern.[1,2] The reasons are many, and among them is the incomplete knowledge researchers in developing countries have regarding the reporting of research.

Early-career researchers generally find it difficult to write research papers. In developed countries, such researchers may receive support from their advisors (who may have mastered the craft of research writing), peers with more experience, and institutional writing centres. In developing countries, these forms of support may be in short supply. Inadequate preparation in research writing can harm the careers of researchers by preventing them from publishing their work, which – compounded with limited funding and time for doing research – may decrease their motivation to conduct further research.

AuthorAID is a concept aimed at supporting developing-country researchers in publishing their work in peer-reviewed journals.[3,4] AuthorAID@INASP is a project run by the International Network for the Availability of Scientific Publications (INASP)[5]. AuthorAID is part of a larger INASP initiative called the Programme for the Enhancement of Research Information (PERii), which also addresses issues such as access to research information, library development, and evidence-informed policy making in developing countries.

The AuthorAID staff at INASP have organised many workshops on research writing in various developing countries since 2007. To expand AuthorAID's training initiative, we considered creating e-learning courses. We started with a pilot phase in which we planned to run a web-based, e-learning course titled "Writing a Research Paper for Publication". The National University of Rwanda (NUR) agreed to be a partner in the pilot phase. Teaching and research faculty at the NUR were encouraged to enrol. The course ran from 3 October to 27 November 2011, with 28 learners and one instructor (the author).

This paper explains the challenges faced, how the course was conducted, and the outcomes.

*Challenges*

E-learning offers the tempting combination of cost-effectiveness and scalability, but making e-learning work can be challenging. For us, the challenges were the following:

1. Low retention rate or high dropout rate, compared to classroom instruction, is a classic problem in e-learning and distance education: attrition or dropout rates are typically 10 to 20% higher (or more) in online courses compared to classroom courses.[6,7,8]

2. In developing countries, e-learning faces additional challenges.[9,10] Barriers to learning online include low-bandwidth or unstable Internet connectivity; lack of computers; and electricity outages. However, National Research and Education Networks (NRENs) in developing countries are improving. Rwanda has such an NREN, and the NUR is part of it.

3. The pilot e-learning course was to be conducted in English, a language that only recently became the medium of educational instruction in Rwanda.

4. The pilot course would be free of cost and not carry any official credit. While the former is meant to be an advantage, combined with the latter it may not be so. Learners who lose interest in the course may drop out because of the lack of both personal investment and tangible benefits. The learners' satisfaction with the online course could be the key to a high retention rate.[11,12] The social presence of the instructor and learners within the online course could also be critical for its success.[13]

*Objective of the pilot course*
Completion rate and learner feedback were to be the indicators of the success of the pilot e-learning course. The objective was to see a completion rate similar to that in AuthorAID workshops (about 90%) and positive feedback from the learners.

**Methods**

*Online learning environment*
Moodle was chosen as the online learning environment for hosting the pilot course. Moodle is free, open-source software[14] that has found acceptance in many universities around the world.

In July 2011, the latest version of Moodle (2.1) was downloaded and installed by AuthorAID's technology partner, the Institute of Learning, Research and Technology (ILRT) in Bristol, UK. Moodle was made available at http://aamood-demo.ilrt.bris.ac.uk/ (this URL may not be permanent). The basic, "default" theme, which has minimal CSS and images, was used so that users on low-bandwidth connections would not face long download times for pages.

*Content*
The foundation for the content was the lectures given at AuthorAID workshops, most of which have been written by Prof Barbara Gastel based on her book on scientific writing.[15]

The course consisted of nine lessons: (1) approaching a writing project; (2) publishing a paper in a journal; (3) the title and authorship; (4) tables and figures; (5) citations and references; (6) the abstract and introduction; (7) the methods section; (8) the results section; and (9) the discussion section. The lessons were spread over five weeks, from 3 October to 6 November 2011. In the last week of the course (7 to 13 November 2011), the learners had to do the assignment. Because the learners were to take the course alongside their teaching and research work, two hours a week was the prescribed time for course work. A screenshot of the course page is shown in Figure 1.

Moodle version 2.1 has 13 types of activities.[16] An activity is something that involves the participation of the learner. In the pilot course, five activities were used: lesson, forum, database, feedback, and glossary.

The lessons were created using the "lesson" activity. They were set up such that a learner could view a lesson only after completing the previous lesson. Two considerations were used in creating the lessons: readability and sustaining learners' interest. To aid readability, (1) each lesson page usually had not more than about 100 words, (2) short sentences, normally 10 to 20 words long, were used, and (3) difficult words were avoided. To sustain the learners' interest, each lesson had a mix of content and thought-provoking questions. Of the 118 pages in all the lessons combined, 74 pages (over 60%) contained such questions. Most questions were multiple-choice, and some involved matching (eg Figure 2).

If a question was answered incorrectly, the learner could try again multiple times. When the learner selected the correct answer, an explanation was given, which began with a smiley to add some levity. For incorrect answers, an explanation was sometimes given; in cases where the explanation could reveal the correct answer, the learner was simply asked to try again. This question behaviour was meant to (1) allow learners to think about questions until they arrived at the correct answer (while not being penalised, because the lessons were not tests) and (2) prevent learners from flying through the lessons without engaging with the content.

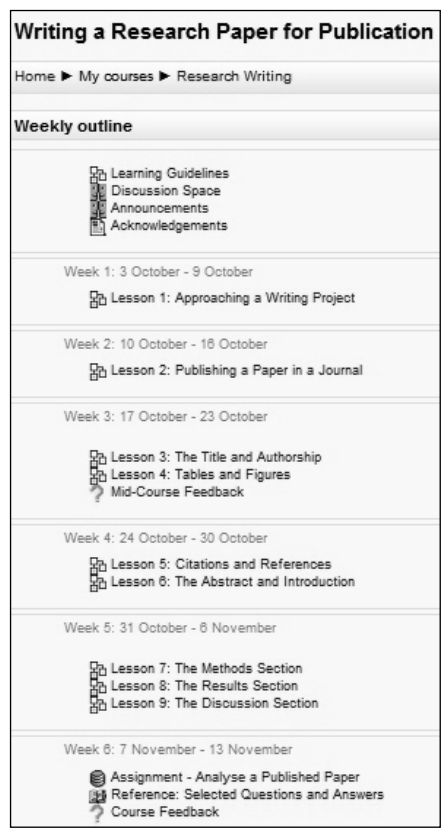

**Figure 1. Screenshot of the course page in Moodle**

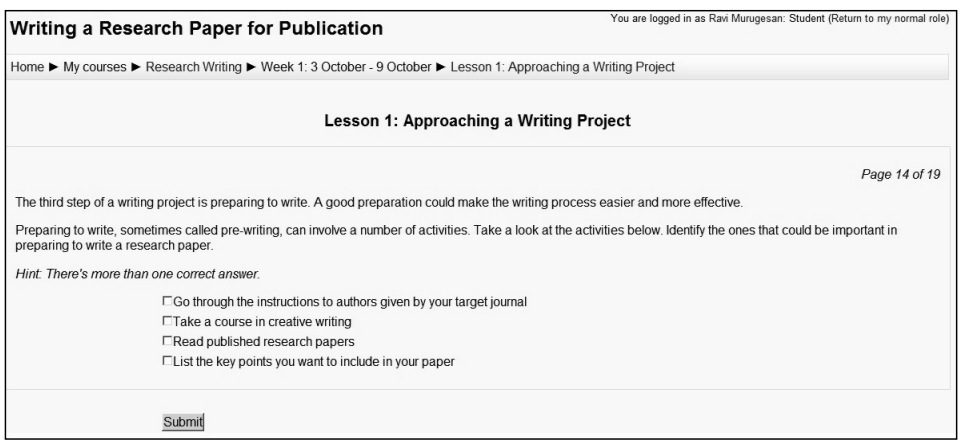

**Figure 2. Screenshot of a page in a lesson**

There was no test or quiz during the course. Because the course aimed to be an online form of the AuthorAID workshop, where there is no test, assessment was not a feature. However, there was an assignment at the end of the course: learners had to analyse published research papers in their field and enter some characteristics of the papers' in a database that all learners could see.

*Getting started*
The success of an e-learning course may depend on making learners aware of the educational technology they are going to use, for example, with a "warm-up" period.[17] Therefore, a preparatory course was set up for potential learners of the pilot course to become familiar with the Moodle environment and the course instructor. The main component of this course was a forum where learners introduced themselves.

Then, at the start of the pilot course, a five-page lesson called "learning guidelines" was provided. This lesson was made in the same style as the actual lessons of the course (with a mix of content and questions). At the end of this lesson, learners were asked to make a post on the course forum (called "discussion space") explaining their learning goals.

The purpose of the preparatory course and the lesson with learning guidelines was to make learners comfortable with the two key elements of the course: the content (lessons) and interaction (forum).

*Enrolment*
We wanted to attract as many learners as possible to test the pilot course, so the course was open to academic staff in all departments, not just scientific researchers. Because the content had more breadth than depth, researchers from other fields could also benefit from the course.

On 10 August 2011, an announcement about the course was sent by e-mail to all academic staff at the NUR. The announcement contained a link for registration on the Moodle site; the site was set up to allow self-registration. At this time, only the preparatory course, not the pilot course, was available. On 23 September 2011, the author announced, in the "news" forum of the preparatory course, that the pilot course was ready and learners could enrol in it. Learners were advised to go through the "learning guidelines" lesson first. On 26 September 2011, a week before the course start date, this announcement was again sent by e-mail to all academic staff at the NUR.

Twenty-five faculty members from NUR enrolled themselves in the preparatory course. Of these, 15 enrolled themselves in the pilot course. Thirteen faculty members joined the pilot course without taking the preparatory course, leading to a total of 28 learners in the pilot course (22 men and six women).

*Moderating the course*
To have a strong online presence, the author maintained contact with the learners in three ways: the discussion forum, where learners could ask questions; the news forum; and e-mail. At the start of the course, the learners were told that they could post their questions on the discussion forum or send them to the author privately by e-mail. The author replied to questions within a day.

The news forum was used for making announcements. The author wrote to the learners about which lesson(s) they had to go through in a particular week, feedback forms being available, how to complete the assignment, and so on. In all, 12 announcements were made.

Whenever a post was made on the discussion and news forums, all the learners in the course received an e-mail alert with a copy of the post. This ensured that everyone was up to date with the posts, even if they were not checking the forums. They could not reply to the post by e-mail; they had to login to the system to do that.

The author used e-mail to motivate learners who were falling behind the course schedule. Once, he wrote an e-mail praising a learner who completed three lessons ahead of schedule.

*Collecting feedback*
Feedback was collected twice during the course: near the middle and at the end. The purpose of the mid-course feedback form was to see how the learners were faring. The learners could fill out both the feedback forms anonymously. They were not required to fill out the forms, but they were given an incentive to fill out the final feedback form: only after doing this did they receive instructions on how to claim their course completion certificate.

**Results**

*Completion rate*
The course was supposed to end on 13 November 2011. On this date, 16 out of 28 learners had completed the course. On 14 November, the author made an announcement in the news forum, informing the learners that there would be a two-week extension for them to complete the course. Nine learners completed the course in the extension period. Therefore, the course completion rate was 89.3% (25 out of 28 learners).

*Feedback*
Eighteen learners filled out the mid-course feedback form, whereas all the 25 learners who completed the course filled out the final feedback form and claimed their certificate. The learners' responses to the multiple-choice questions in these forms are summarised in Tables 1 and 2. Some questions did not have any neutral options; these are marked with a dash in the "neutral" column.

After the course, the author e-mailed the three learners who had not completed the course. They were asked why they could not complete the course and whether they would be interested in taking it in the future. Two of them replied; both mentioned personal reasons that had made them indisposed and expressed interest in taking the course later. One of them started going through the lessons immediately after responding to the e-mail and completed the course within two weeks.

| Question | Positive | Negative | Neutral |
|---|---|---|---|
| 1. How has your learning experience been so far? | 100% | 0% | - |
| 2. Each lesson has a number of questions. Do you like this format? | 83% | 0% | 17% |
| 3. How do you find the level of English used in the lessons? | 100% | 0% | - |
| 4. How do you find the course schedule? | 89% | 11% | - |
| 5. Has your environment been suitable for your study? | 78% | 0% | 22% |
| 6. Are you getting enough support from the course facilitator? | 89% | 0% | 11% |

**Table 1. Mid-course feedback: Summary of responses to the multiple-choice questions from 18 learners**

| Question | Positive | Negative | Neutral |
|---|---|---|---|
| 1. What was your knowledge of research writing just before you started this course? | 64% | 36% | - |
| 2. What do you think is your knowledge of research writing now, after completing the course? | 100% | 0% | - |
| 3. Do you feel you are now ready to write a research paper for publication? | 96% | 0% | 4% |
| 4. Do you think you can pass on the knowledge you gained from this course to others? | 96% | 0% | 4% |
| 5. If you had any questions during the course, did you feel comfortable asking those questions? | 76% | 0% | 24% |
| 6. If you asked any questions in the course - either on the Discussion Space or by e-mail, were they answered clearly and promptly? | 88% | 4% | 8% |
| 7. Would you have liked to work with the other participants in the course to complete exercises? | 52% | 24% | 24% |
| 8. Rate the quality of the interactive lessons. | 100% | 0% | 0% |
| 9. Rate the quality of the summary documents provided at the end of the lessons. | 76% | 4% | 20% |
| 10. Did the assignment in the course (analysing one or more published research papers) help you learn more about research writing? | 100% | 0% | - |

**Table 2. Final feedback: Summary of responses to the multiple-choice questions from the 25 learners who completed the course**

*Interaction*
Nineteen of the 28 learners made an introductory post on the discussion forum. Nine learners used the discussion forum to ask one or more questions during the course. Three learners e-mailed the author during the course with questions.

In the feedback form was this question: If you had any questions during the course, did you feel comfortable asking those questions? Nineteen learners answered "yes", and six said "I did not have any questions". No one marked "no", which indicates that the learners were either comfortable asking questions or did not have questions. However, only 12 learners asked questions, although 19 learners made an introductory post. Perhaps this question was interpreted as "were you comfortable making a post on the discussion forum".

Towards the end of the course, the glossary module was used to present selected questions that learners had asked during the course and the answers.

**Discussion**
E-learning was new for most of the learners. Through the feedback form at the end of the course, it was found that 19 of the 25 learners (76%) who completed the course had not taken an e-learning course before. First-time e-learners find it especially difficult to complete e-learning courses.[18] Yet, the course completion rate was high: at 89%, it is similar to the completion rate seen in AuthorAID workshops. Given the volume of academic literature devoted to the topic of attrition in e-learning, this high completion rate is noteworthy.

Of the 16 learners who completed the course by the original end date, 13 did so more than a week in advance. Only three learners completed the course in the last week. The low completion rate in the last week could have been because the author was away that week, conducting a workshop in Ethiopia. Reminders or encouraging messages could not be sent to the learners that week. Only one announcement was made in the last week (out of 12 made throughout the course), and that message was about the author being away. The low online presence of the author in that week could have had an impact on the learners' progress in the course. As soon as the author returned and made a post about the two-week extension period, the remaining learners made steady progress. Eight of the nine learners who completed the course in this period did so in the first week itself.

Feedback from the learners (Tables 1 and 2) is positive and encouraging. In addition to the multiple-choice questions in the feedback forms, there were a number of short-answer questions, for example, "Please specify what you think you will do differently, including any specific plans on sharing of skills/knowledge, as a result of taking the course". The learners' responses reveal that they found the course very useful. Based on the feedback, the main point that requires further attention is enabling group work in the course. This was not a feature of the pilot course, in contrast to AuthorAID workshops that have group activities in which participants work on their own research writing. Such activities can be part of e-learning courses in Moodle, for example, the "workshop" module is meant for peer assessment.

As for interaction, 42% of the learners (12 out of 28) asked questions during the course, while 68% made at least an introductory post. The level of interaction was not high; however, no learner marked "no" to question number 5 in Table 2, so at least the learners were not uncomfortable asking any questions they did have. Two of the three learners who asked questions by e-mail presented a total of seven questions. In contrast, the nine learners who asked questions over the discussion forum usually asked one question each. This could indicate that learners prefer to e-mail the instructor directly when they have many questions.

## Conclusion

The objectives of the pilot course were achieved: the completion rate was similar to that in AuthorAID classroom workshops, and the learners gave positive feedback. Therefore, e-learning is viable for AuthorAID's training objectives, and it may be so for others involved in teaching research or scientific writing to researchers in developing countries.

The success of the course can be attributed to the following: (1) providing a preparatory course and learning guidelines before starting the course at a gentle pace; (2) presenting content that sustained the learners' interest and was appropriate for their language level; and (3) keeping in touch with the learners throughout the course by answering questions promptly, writing about current and upcoming topics, and paying attention to those falling behind.

A training programme is successful if the learners accomplish something by applying their learning. Workshops (and in the future online courses) run by AuthorAID aim to equip researchers with the knowledge and skill to publish in peer-reviewed journals. AuthorAID workshops have indeed led to increased publications, and it is hoped that online courses will too.

## Acknowledgements
Prof Verdiana Masanja and Mr Gilbert Munyemana at the National University of Rwanda provided encouragement, support, and advice for running the course. Ms Julie Walker at INASP made strategic contributions, Prof Barbara Gastel reviewed the initial lessons in the course, and Ms Sara Gwynn critiqued this manuscript. The author is grateful to Penny Hubbard for her comments.